\documentclass[twocolumn,aps,prc,showpacs,superscriptaddress,preprintnumbers,floatfix,nofootinbib]{revtex4}
\usepackage{epsfig,graphics}
\usepackage{graphicx}% Include figure files
\usepackage{dcolumn}% Align table columns on decimal point
\usepackage{bm}% bold math
\usepackage{amsmath}
\usepackage[usenames]{color}
\usepackage{ulem} %% for strike-through

\usepackage{CJK}

\begin{document}
%\begin{CJK*}{GBK}{song}

\title{Thermodynamic properties and shear viscosity over entropy density ratio of nuclear fireball in a quantum-molecular dynamics model}

\author{C. L. Zhou}
\affiliation{Shanghai Institute of Applied Physics, Chinese
Academy of Sciences, Shanghai 201800, China}
\affiliation{University  of Chinese Academy of Sciences, Beijing 100049, China}

\author{ Y. G.  Ma\footnote{Author to whom all correspondence should be addressed:
ygma@sinap.ac.cn}}
\affiliation{Shanghai Institute of Applied Physics, Chinese
Academy of Sciences, Shanghai 201800, China}

\author{ D. Q. Fang}
\affiliation{Shanghai Institute of Applied Physics, Chinese
Academy of Sciences, Shanghai 201800, China}

\author{ G. Q.  Zhang}
\affiliation{Shanghai Institute of Applied Physics, Chinese
Academy of Sciences, Shanghai 201800, China}

\date{\today}

\begin{abstract}
Thermodynamic and transport properties of nuclear fireball  created in the central region of heavy-ion collisions below 400 MeV/nucleon are investigated within the isospin-dependent quantum molecular dynamic (IQMD) model. These properties including the density, temperature, chemical potential, entropy density ($s$) and shear viscosity ($\eta$), are calculated by a generalized hot Thomas Fermi formulism and a parameterized function, which was developed by Danielewicz. As the collision goes on, a transient
minimal $\eta/s=5/4\pi-10/4\pi$ occurs in the largest compression stage. Besides, the relationship of $\eta/s$ to temperature ($T$) in the freeze-out stage displays a local minimum which is about 9-20 times $1/4\pi$ around $T$ = 8-12 MeV, which can be  argued as indicative of a liquid gas phase transition. In addition, the influences of nucleon-nucleon (NN) cross section ($\sigma_{NN}$) and symmetry energy coefficient ($C_{s}$) are also discussed, and it is found that the results are sensitive to $\sigma_{NN}$ but not to $C_{s}$.
\end{abstract}

\maketitle

\section{\label{sec:level1}INTRODUCTION}

In the past decades, extensive experimental and theoretical
efforts have been devoted to search for the nuclear liquid-gas
phase  transition (LGPT) in  intermediate energy heavy-ion
collisions (HIC)
\cite{Gupta,Bona,Bor,Poc,Ma-plb,Nato,Ma-NST,Gil,Ell,Ma-PRL,Gross1990,Bondorf1995}.
Many probes have been suggested for the onset of nuclear LGPT. For
instance, the fragment size distribution \cite{Fisher} and its
rank distribution \cite{Ma-PRL}, the largest fluctuation of the
heaviest fragment \cite{fluct}, caloric curve \cite{Poc,Nato},
bimodality \cite{Lopez} etc. In addition, it has been observed
that the ratio of shear viscosity to entropy density ($\eta/s$)
reaches its local minimum at the transition temperature for a wide
class of systems. For instance, empirical observation of the
temperature or incident energy dependence of the shear viscosity
to entropy density ratio for H$_2$O, He and Ne2 exhibits a minimum
in the vicinity of the critical point for phase transition
\cite{Csernai2006}. And a lower bound of $\eta/s > 1/4\pi$
obtained by Kovtun-Son-Starinets (KSS) for infinitely coupled
super-symmetric Yang-Mills gauge theory based on the AdS/CFT
duality conjecture, is speculated to be valid universally
\cite{Kovtun2005,Policastro2001}. In ultra-relativistic HIC
\cite{Demir2009,Lacey2007,Chen,Kapu2008,Maj2007}, people
have used the ratio of shear viscosity to entropy density to study
the quark-gluon plasma phase and the extracted value of $\eta/s$
seems close to the KSS bound ($1/4\pi$).

So far there are many interesting investigations on the  ratio of
$\eta/s$, but it is still rare to study the behavior of $\eta/s$
during the heavy-ion collision at intermediate energies \cite{Shi2003,Pal2010,Auerbach2009,Li2011}.
Furthermore, the influences of nucleon-nucleon cross section  and
nuclear symmetry energy are less discussed.

In this work we use a microscopic transport model known as the
isospin-dependent  quantum dynamics  model \cite{Ma2001} to
simulate Au+Au central collisions. In order to study the effect of
nucleon-nucleon cross section, 0.5 times and 1.5 times normal
nucleon-nucleon  cross section are also used in the simulation. On
the other hand for the symmetry energy which is important for asymmetric nuclear matter as well
as nuclear astrophysics \cite{asy1,asy2,asy3}, different symmetry
energy parameters are employed, namely 15 MeV, 25 MeV and 35 MeV. The
generalized hot Thomas Fermi formalism (GHTFF)
\cite{khoa1992a,khoa1992b,puri1992} and the transport formula
\cite{Danielewicz1984} are employed , respectively,  to extract
thermodynamic and transport properties of the nuclear fireball
which is located in the central region with a moderate volume. Then different correlations between the extracted thermal and transport properties are discussed and a good agreement
with our previous calculations is found \cite{zhoucl2012}.  Furthermore the multiplicity of intermediate mass fragments (IMFs) is also checked as a signal of liquid gas phase transition \cite{Ma1995,Peaslee,Ogilvie,Tsang} to verify the calculated result.

The paper is organized as follows. Section 2 provides a brief
introduction  for the IQMD model, GHTFF as well as transport formula for shear viscosity. In
Section 3 we present the calculation results and discussions,
where the time evolution of thermodynamic quantities and shear
viscosity over entropy density are focused. Finally a summary and outlook is given.
%%%%%%%%%%%%%%%%%%%%%%%%%%%%%%%%%%%%%%%%%%%%%%%%%%%%%%%%%%%%%%%%%%%%

\section{\label{sec:level2} model and formulism}

\subsection{Quantum Molecular Dynamics  Model}

The quantum molecular dynamics (QMD)
\cite{Hartnack1998,Aichelin1991} model  approach is a many-body
theory which describes heavy ion collisions from intermediate to
relativistic energy. The isospin-dependent quantum molecular
(IQMD) \cite{Hartnack1989,Hartnack1998} model is based on the QMD
model, including the isospin effects and Pauli blocking. Each
nucleon in the colliding system is described as a Gaussian wave
packet, i.e.
\begin{eqnarray}
\psi_i({\bf p_i},{\bf r_i},t)&=&\frac{1}{(2\pi
L)^{3/4}} \exp [ \frac{i}{\hbar} {\bf p}_i(t)\cdot{\bf
r}. \nonumber \\
&& -\frac{({\bf r}-{\bf r}_i(t))^2}{4L}].
\end{eqnarray}
Here ${\bf r}_{i}(t)$ and ${\bf p}_{i}(t)$ are the mean position
and mean momentum,  and the Gaussian width has the fixed value
$L=2.16 fm^{2}$ for Au + Au system. The centers of these Gaussian
wave packets propagate in coordinate (${\bf R}$) and momentum
(${\bf P}$) space according to the classical equations of motion:
\begin{equation}
\dot{{\bf p}}_{i}=- \frac{\partial \langle {\bf H}
\rangle}{\partial {\bf r}_i}; ~~~~~\dot{{\bf r}}_i=\frac{\partial
\langle {\bf H} \rangle}{\partial {\bf p}_i},
\end{equation}
where $\langle {\bf H} \rangle$ is the Hamiltonian of the system.

The Wigner distribution function for a single nucleon density in phase space is given by
 \begin{eqnarray}
 f_{i}({\bf r},{\bf p},t) &=& \frac{1}{(\pi\hbar)^3} \exp \left[ \frac{-({\bf r}-{\bf r}_i(t))^2}{2L} \right] \nonumber \\
 & & \exp \left[ \frac{-2L({\bf p}-{\bf p}_i(t))^2}{\hbar^2}\right].
\end{eqnarray}

The mean field in IQMD model is written as
\begin{equation}
U(\rho) = U_{\rm Sky} + U_{\rm Coul}  + U_{\rm Yuk} + U_{\rm sym},
\end{equation}
 where $U_{\rm Sky}$, $U_{\rm Coul}$, $U_{\rm Yuk}$, and $U_{\rm
sym}$  represents the Skyrme potential, the Coulomb potential, the
Yukawa potential and the symmetry potential interaction,
respectively \cite{Aichelin1991}.
The Skyrme potential is
\begin{equation}
U_{\rm Sky} = \alpha(\rho/\rho_{0}) + \beta{(\rho/\rho_{0})}^{\gamma},
\end{equation}
 where $\rho_{0}$ = 0.16
{fm}$^{-3}$ and $\rho$ is the nuclear density. In the present work, the parameters
$\alpha =-356$ MeV, $\beta =303$ MeV, and $\gamma = 7/6$,
correspond to a soft EOS, are used. $U^{\rm Yuk}$ is
a long-range interaction (surface) potential, and takes the
following form
\begin{eqnarray}
 \label{v_yuk}
&U^{Yuk} & =  ({V_y}/{2}) \sum_{i \neq j}{exp(Lm^2)}/{r_{ij}} \nonumber \\
&\cdot & [exp(mr_{ij})erfc(\sqrt{L}m-{r_{ij}}/{\sqrt{4L}}) \nonumber \\
&-& exp(mr_{ij})erfc(\sqrt{L}m+{r_{ij}}/{\sqrt{4L}})],
\end{eqnarray}
 with $V_y =0.0074$GeV, $m$ = 1.25{fm}$^{-1}$,  $L = 2.16$ fm$^{2}$, and
$r_{ij}$ is the relative distance between two nucleons. The
symmetry potential is $U_{sym} = C_{s}\frac{\rho_n
-\rho_p}{\rho_0}$, where $\rho_n$, $\rho_p$, and $\rho_0$ are the
neutron,  proton and nucleon densities, respectively. $C_{s}$ is
the  symmetry energy coefficient, and three different values of
15, 25 and 35 MeV are taken in order to study its influence on the ratio
of $\eta/s$.

Furthermore the isospin degree has entered into the cross sections,
which is similar to the parametrization of VerWest and Arndt, see
Ref.~\cite{bass1995}. The cross section for the neutron-neutron
collisions is assumed to be equal to the proton-proton cross
sections. In order to study the effect of cross section on the
ratio of $\eta/s$, the nucleon-nucleon cross section is multiplied
by a coefficient $C_{\sigma}$. Three different situations are
considered, namely  $C_{\sigma}$  equals 0.5, 1.0 and 1.5
respectively . In a practical
viewpoint, a smaller $C_\sigma$ seems suitable to describe HIC, which
was proposed in the previous work  Ref. \cite{Danielewicz2009} 
\begin{equation}
\sigma_{NN}=C_{\sigma}\sigma_{NN}^{free}.
\end{equation}

From Eq. (3) one obtains the matter density of coordinate space by the sum over all the nucleons, namely
\begin{eqnarray}
\label{eq-rhom}
 \rho({\bf r},t) &= &\sum_{j=1}^{A_{T}+A_{P}}\rho_{j}({\bf r},t) \nonumber \\  && = \sum_{j=1}^{A_{T}+A_{P}} \frac{1}{(2{\pi}L)^{3/2}} \exp\frac{-({\bf r}-{\bf r}_i(t))^2}{2L}.
\end{eqnarray}

The kinetic energy density in coordinates space could also be calculated from Eq.(4) by
\begin{eqnarray}
\label{eq-taom}
\rho_{K}({\bf r},t)= \sum_{j=1}^{A_{T}+A_{P}}\frac{{{\bf P}_j(t)}^2}{2m }\rho_{j}({\bf r},t).
\end{eqnarray}

\subsection{The Generalized Hot Thomas-Fermi Formalism}

Thermodynamical properties of hot nuclear matter formed  in
heavy ion collisions, e.g. temperature and entropy density, can be
extracted by using the approach developed by Faessler and
collaborators
\cite{khoa1992a,khoa1992b,puri1992,barranco1981,rashdan1987}. In
this approach one starts from a microscopic picture of two
interpenetrating pieces of nuclear matter and deduces the thermal
quantities from the matter density and kinetic energy density
obtained during the collisions. In this paper, the extraction of
thermal properties of the hot nuclear matter is done in two steps.
First, based on the IQMD simulation, one could calculate the
nuclear matter and kinetic energy densities at each point in
coordinate space at every time step. Second, by employing the hot
Thomas-Fermi formalism, we could obtain the corresponding thermal
properties for every set of nuclear matter density and nuclear
kinetic energy density\cite{khoa1992a,khoa1992b}.
In GHTFF, the momentum distribution in cylindrical coordinates $k_{r},k_{z}$ can be written as
\begin{multline}
n(K)=
\left\{
\begin{array}{ll}
n1(K)= \\
n2(K)= \nonumber \\
\end{array}
\right.\\
\left\{
\begin{array}{ll}
(1+exp[\hbar^{2}(k_{r}^{2}+k_{z}^{2})/2mT-\mu_{1}^{'}])^{-1}, & k_{z} < k_{0} \\
(1+exp[\hbar^{2}(k_{r}^{2}+(k_{z}-k_{R})^{2})/2mT-\mu_{2}^{'}])^{-1}, & k_{z} > k_{0} \nonumber
\end{array}
\right.
\end{multline}
with $\mu_{i}^{'}=\mu_{i}/T$ is the reduced chemical potential, $k_{0}=[k_{R}^{2}-2mT(\mu_{1}^{'}-\mu_{2}^{'})]/2k_{R}$, $k_{R}$ is the relative momentum between the
projectile (index 1) and target (index 2).
The local nuclear matter density $\rho_{i}$ is expressed as
\begin{eqnarray}
\label{eq-rho}
\rho_{i}&=&\frac{1}{2}\rho_{0}(\mu_{i}^{'})+\frac{1}{2\pi^2}
(\frac{2mT}{\hbar^2})^{3/2} \nonumber \\ &&\times
[f(\mu_{i}^{'},K_{0i})+J_{1/2}(\mu_{i}^{'},K_{0i}^2)],
\end{eqnarray}
where $K_{01}=\frac{\hbar K_{0}}{\sqrt{2mT}}, K_{02}=K_{R}-K_{01}$
with $K_{r}=\frac{\hbar k_{R}}{\sqrt{2mT}}$, and $J_{n}(\mu^{'})=J_{n}(\mu^{'},\infty)$ is the Fermi integrals, i.e.
\begin{eqnarray}
J_{n}(\mu^{'},z)=\int_{0}^{z}\frac{x^{n}dx}{1+\exp(x-\mu^{'})},\nonumber
\end{eqnarray}
$f(\mu_{i}^{'},K_{0i})=K_{0i}ln[1+\exp(\mu_{i}^{'}-K_{0i}^2)]$.\\
The local kinetic energy density $\epsilon=\frac{\hbar^2\tau_{i}}{2m}$, where $\tau_{i}$ reads
\begin{eqnarray}
\label{eq-tao}
\tau_{i}&=&\frac{1}{2}\tau_{0}(\mu_{i}^{'})+\frac{1}{2\pi^2}(\frac{2mT}{\hbar^2})^{5/2}
\nonumber
\\&&\times[\frac{1}{3}K_{0i}^2f(\mu_{i}^{'},K_{0i})+\frac{1}{3}J_{1/2}(\mu_{i}^{'},K_{0i}^2)
\nonumber \\ &&+\int_{0}^{K_{0i}}J_{1}(\mu_{i}^{'}-x^2)dx
]+\Delta\tau_{i}(\mu_{i}^{'}).
\end{eqnarray}
And the entropy density $s_{i}$ is written as
\begin{eqnarray}
\label{eq-si}
s_{i}&=&\frac{1}{2}s_{0}(\mu_{i}^{'})+\frac{1}{2\pi^2}(\frac{2mT}{\hbar^2})^{3/2} \nonumber
\\&&\times[(\frac{1}{3}K_{0i}^2-\mu_{i}^{'})f(\mu_{i}^{'},K_{0i})+\frac{1}{3}J_{1/2}(\mu_{i}^{'},K_{0i}^2)\nonumber
\\&&-\mu_{i}^{'}J_{1/2}(\mu_{i}^{'},K_{0i}^2)+2\int_{0}^{K_{0i}}J_{1}(\mu_{i}^{'}-x^2)dx].
\end{eqnarray}
Here $i=1,2$ represents the projectile and target, and $\Delta\tau_{1}(\mu_{1}^{'})=0$,\\
\begin{eqnarray}
\Delta\tau_{2}(\mu_{2}^{'})&=&\frac{1}{2\pi^2}(\frac{2mT}{\hbar^2})^{5/2}K_{R}\nonumber \\
&&\times[J_{1}(\mu_{2}^{'})-J_{1}(\mu_{2}^{'},K_{02}^2)-K_{02}f(\mu_{2}^{'},K_{02})]\nonumber \\&&+k_{R}^2\rho_{2}(\mu_{2}^{'}),\nonumber
\end{eqnarray}
From the Eq.(\ref{eq-rho},\ref{eq-tao},\ref{eq-si}), one can obtain the thermal properties by inversion in principle. But such an inversion procedure is practically not feasible due to the complexity of the equations. Therefore, a more practical way is chosen to obtain the thermal properties. First, we generate all reasonable combinations $T$, $K_{R}$ and $\mu_{i}^{'}$, which ranging from 0-100MeV, 0-5$fm^{-1}$ and 0-2, respectively. Then the corresponding $\rho_{i},\tau_{i},s_{i}$ could be obtained. Second, from the extracted $\rho_{i},\tau_{i}$ in the central region at each time step during the evolution of collision, $T$, $K_{R}$ and $\mu_{i}^{'}$ are obtained from the calculations in the first step. Third, the entropy density is calculated according to Eq.\ref{eq-si}. One should pay attention that all the values displayed in the following pictures are the average one in the central region.

\subsection{Shear Viscosity Formalism}

\begin{figure}
\includegraphics[width=10cm]{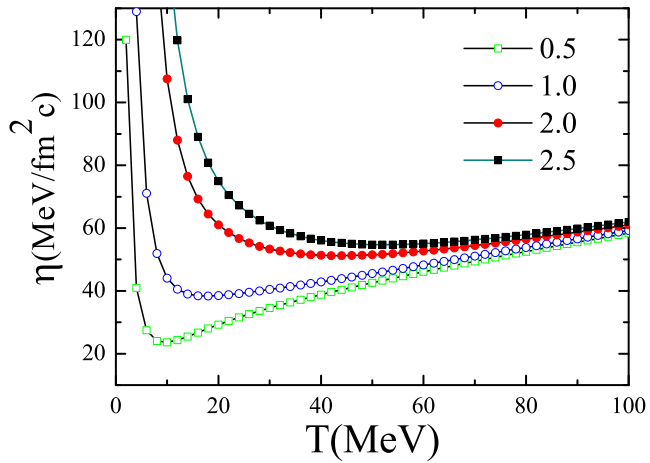}
\vspace{-0.1truein} \caption{\footnotesize (Color online)  Shear viscosity of nuclear matter
as a function of $\rho/\rho_{0}$ and $T$ with the Eq.(\ref{Dani-eqn}). Different colors represent different
$\rho/\rho_{0}$, which are illustrated in the inset.
}\label{Fig_eta-eqn}
\end{figure}

For largely equilibrated systems, fluxes of macro quantities,
leading to dissipation, are proportional to gradients within the
system. The shear viscosity denoted as $\eta$ is the coefficient
of proportionality between anisotropy of momentum-flux tensor,
including dissipation and velocity gradients \cite{kubo}. In the Boltzmann
statistical limit the shear viscosity corresponds to the first
order Chapman-Enskog coefficients. In
Ref.~\cite{Danielewicz1984,Danielewicz2009} the nuclear shear
viscosity for normal N-N cross section, has been derived from the microscopic Boltzmann-Uehling-Ulenbeck equation and
can be parameterized as a function of density $\rho$  and
temperature $T$:
\begin{eqnarray}
\eta(\frac{\rho}{\rho_{0}},T) &=& \frac{1700}{T^2}(\frac{\rho}{\rho_{0}})^{2}+\frac{22}{1+T^{2}10^{-3}}(\frac{\rho}{\rho_{0}})^{0.7}  \nonumber \\
&& + \frac{5.8\sqrt{T}}{1+160T^{-2}},
\label{Dani-eqn}
\end{eqnarray}
where $\eta$ is in MeV/fm$^2 c$, $T$ in MeV, and $\rho_{0}$=0.168fm$^{-3}$.
Figure \ref{Fig_eta-eqn} shows $\eta$ as a function of $T$ and
$\rho/\rho_{0}$. One can see  that $\eta$ exhibits a very distinct
minimum when nuclear matter density is less than normal nuclear
density. And as the density increases, the transition temperature
also get larger, e.g. for normal density the transition temperature
locates around 10 MeV, but for 2.5 times normal density it is almost
50 MeV. This conclusion is coincident with macroscopic result.
And in the case of scaled N-N cross section, the shear viscosity just only needs to be scaled by $\frac{1}{C_{\sigma}}$, ie.
\begin{eqnarray}
\eta(\frac{\rho}{\rho_{0}},T,C_{\sigma}) &=& \frac{\eta(\frac{\rho}{\rho_{0}},T)}{C_{\sigma}}
\label{Dani-eqn-cross}
\end{eqnarray}
From Eq.\ref{Dani-eqn-cross} we can see that the shear viscosity is very sensitive to the N-N cross section, the larger the cross section is, the smaller the shear viscosity.
And it is intuitive that large N-N cross section makes the transport of particle momentum much difficult.
Since the equilibrium of the considered nuclear matter at the very
starting stage is not reached, the calculated shear viscosity should be
considered as the transport properties of largely equilibrated
nuclear matter with the same density and kinetic energy density.

\section{\label{sec:level3} calculation and discussion}

In present work, we simulate head-on collision of Au + Au at different beam energies.
The reason why we choose the central collision is that the participant zone is the maximal
and the matter is more nuclear liquid-like during the early time evolution of collision.

\subsection{Time evolution of density profile in reaction plane }
%%%%%%%%%%%%%%%%%%%%%%%%%%%%%%%%%%%%%%%%%%%%%%%%
\begin{figure*}[htbp]
\hspace{-1.23truein}
\hspace{1.7cm}
\includegraphics[width=18cm]{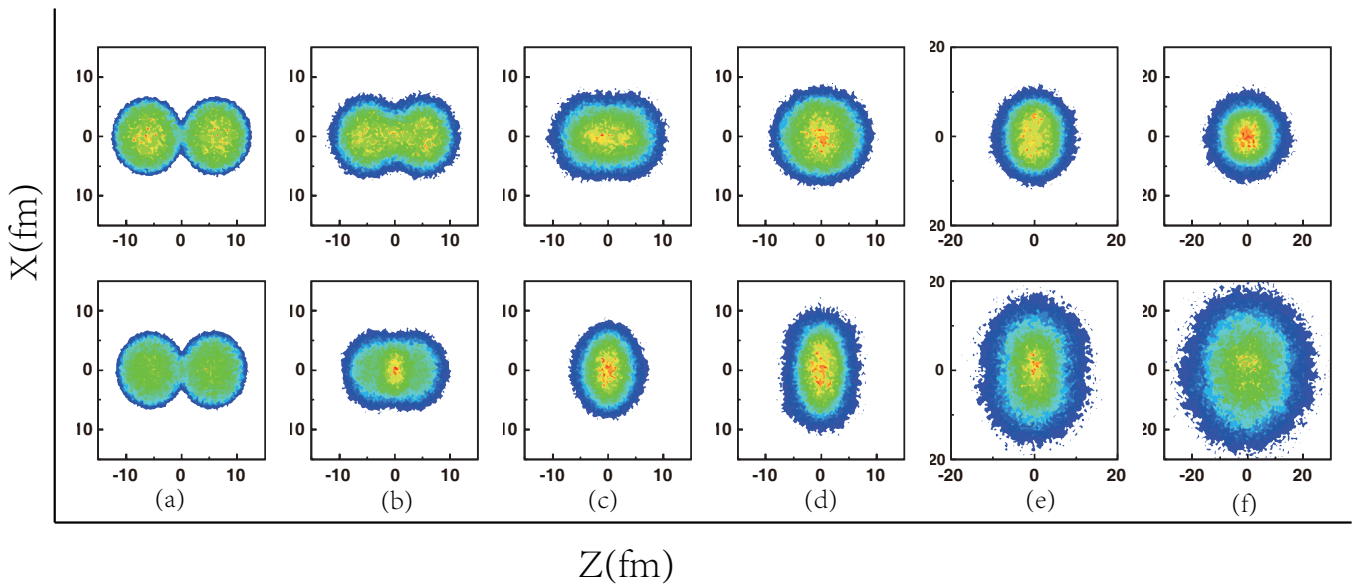}
%\vspace{-1.2truein} 
\caption{\footnotesize (Color online) X-Z
density profile in  different time step for Au+Au head-on
collisions at 50 MeV/nucleon (upper panels) and 200 MeV/nucleon
(lower panels), respectively. From left to right
panels, time step is 2, 10, 20, 30, 50 and 80 fm/c, respectively.
}\label{Fig_xz}
\end{figure*}
%%%%%%%%%%%%%%%%%%%%%%%%%%%%%%%%%%%%%%%%%%%%%%%%%
Fig.~\ref{Fig_xz} shows the evolution of density profile in X-Z
plane  for  Au+Au head-on collisions at 50 MeV/nucleon and 200
MeV/nucleon at 2, 10, 20, 30, 50 and 80 fm/c. The
zero point of time is set at the initial contact between project
and target (the first left panel).  With the collision goes on,
the system overlaps and seems more isotropic in phase space. In
order to calculate the thermodynamic quantities in different time
steps,  we select the central sphere with radius $r$=5 fm, which  defines a
volume of nuclear fireball in this paper.

\subsection{Time evolution of thermodynamic variables}

\begin{figure}[hbp]
\includegraphics[width=9cm]{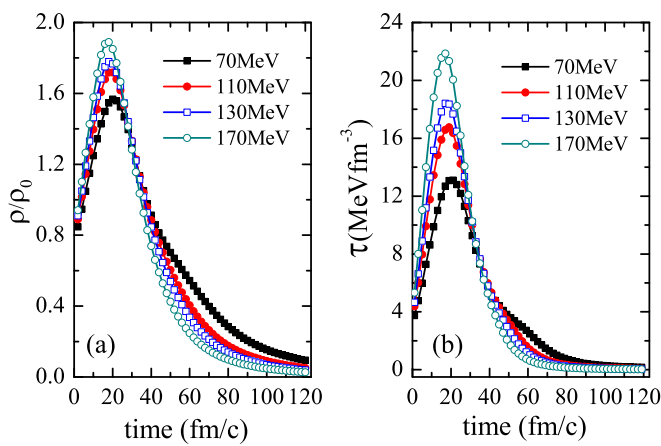}
\vspace{-0.1truein} \caption{\footnotesize (Color online) Time evolution of mean matter density (a) and kinetic energy density (b) at different beam energies.
}\label{Fig_rho-time}
\end{figure}

The time evolution of the average nuclear matter density (panel
(a)) and  kinetic energy density (panel (b)) in the central
region, with $C_{\sigma}=1.0$, is showed in
Fig.~\ref{Fig_rho-time}. It is interesting that the maximum
density reached is about $1.5\rho_{0}$ to $2.0\rho_{0}$ and the
maximum kinetic energy density is 10 MeV fm$^{-3}$ to 25 MeV
fm$^{-3}$ for the energy displayed in the picture. Along the time
scale of the collision one can see that both $\rho/\rho_{0}$ and
$\tau$ are reaching their maxima at about 20 fm/c and at a bit earlier time
for higher energy. After the compression stage the matter starts
to expand and some of them will escape from the central region, mainly in the
transverse plane, the matter density drops to very small values
and the central region is cooled down. In general the warm and dense
nuclear matter survives much longer when the incident energy is low.
At about 80 fm/c the hot and dense matter disappears.

%%%%%%%%%%%%%%%%%%%%%%%%%%%%%%%%%%%%%%%%%%%%%%%%%%%%%%%
\begin{figure}
\includegraphics[width=9cm]{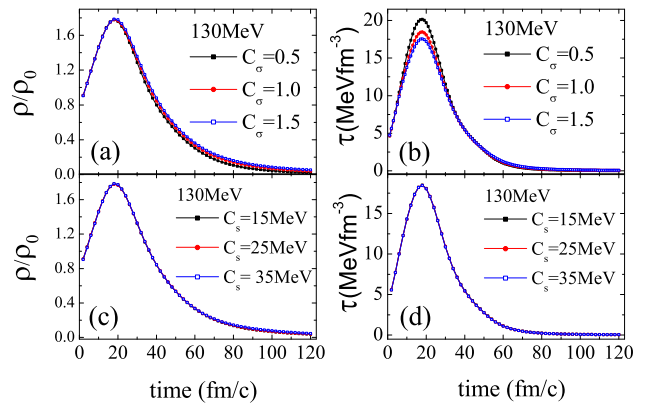}
\vspace{-0.1truein} \caption{\footnotesize (Color online) Time evolution of mean nuclear matter density ( (a) and (c)), and the kinetic energy density ((b) and (d))
 in the central region defined as a sphere with radius equals 5 fm of  head-on Au+Au collisions at 130 MeV/nucleon. Different nucleon-nucleon cross section ((a) and (b)) and symmetry energy parameter ((c) and (d)) are used.
}\label{Fig_130mev-time}
\end{figure}

%%%%%%%%%%%%%%%%%%%%%%%%%%%%%%%%%%%%%%%%%%%%%%%%%%%%%%%
Time evolution of matter density and kinetic energy density  are
shown in Fig.~\ref{Fig_130mev-time}, when $C_{\sigma}$ and $C_{s}$
are set by different values . The upper panels demonstrate the
different cross section situation ($C_{\sigma}\in[0.5,1.0,1.5]$),
on the other hand the bottom panels are for different symmetry
energy. In panels (a) and (b), it is easy to find that the nuclear
matter density and kinetic energy density are different from each
other when $C_{\sigma}$ is different. We found that there is no
difference for the nuclear matter density during  compression
stage. But when the system starts to expand, the larger nucleon-nucleon cross
section  makes the dense matter stay longer. In contrast with
the behavior of density around the maximum compression stage,
more distinction for the kinetic energy density is exhibited. It
shows that the smaller the nucleon-nucleon  cross section, the larger the kinetic
energy density. But these curves almost overlap each other after
40 fm/c, it may  be understood that the longitudinal
energy enters the central region more easily when the nucleon-nucleon cross
section is small. It should be noted that the large kinetic energy
density dose not mean higher temperature, since the kinetic energy
is not calculated in the center of mass frame, detailed
information can be found in
Refs.~\cite{khoa1992a,khoa1992b,puri1992,barranco1981,rashdan1987}.
In addition, the extracted density and kinetic energy density show
insensitivity to the symmetry energy as depicted in panel (c) and
(d). The curves are overlapped with each other for the whole
process.
As has been discussed in the previous paragraph, this leads the nuclear
matter to the same thermal properties. So the extracted thermal
properties based upon the hot Thomas-Fermi formulism keep exactly
the same with each other. So in this paper we just investigate the
nucleon-nucleon cross section effect on thermodynamic and transport quantities in the following texts.

%%%%%%%%%%%%%%%%%%%%%%%%%%%%%%%%%%%%%%%%%%%%%%%%%%%%%%%
%%%%%%%%%%%%%%%%%%%%%%%%%%%%%%%%%%%%%%%%%%%%%%%%%%%%%%%

\begin{figure}
\includegraphics[width=9cm]{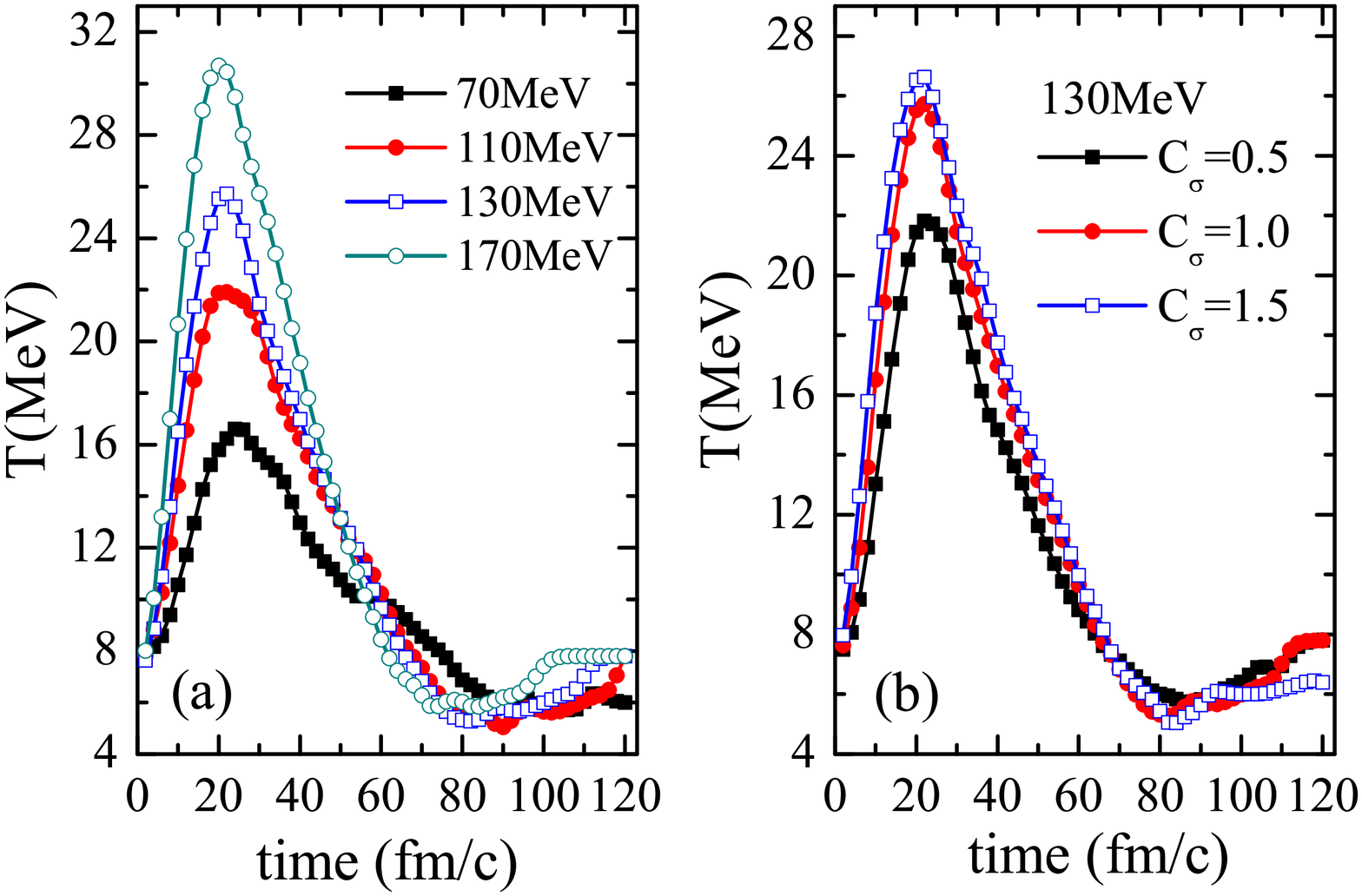}
\vspace{-0.1truein} \caption{\footnotesize (Color online) Time evolution of temperature inside the central region at normal nucleon-nucleon cross section at different incident energies (a), or at 130 MeV/nucleon but with different  $\sigma_{NN}$ (b). The incident energies and N-N cross section parameters are illustrated in the inset.
}\label{Fig_temper-merge}
\end{figure}

Time evolution of temperature is plotted in
Fig.~\ref{Fig_temper-merge}.  Fig.~\ref{Fig_temper-merge}(a) shows the time evolution
of temperature with $C_{\sigma}$=1.0 at different incident
energies. Fig.~\ref{Fig_temper-merge}(b) shows the time evolution of temperature when
the incident energy is fixed at 130 MeV/nucleon but with different
cross sections. The following pictures are arranged with the
same mode, i.e panel (a) represents a constant cross section
$C_{\sigma=1.0}$ but at  different incident energies; panel (b)
means a constant incident energy at 130 MeV/nucleon with different
cross sections. For a given beam energy, temperature increases at
first, then reaches a local maximum about 20 fm/c and decreases
till a saturated value at about 80 fm/c. The higher the incident
energy, the larger the maximum value.  The corresponding time at
maximum  value is a little earlier than that for the density and
kinetic density.
In panel (b), it is found that the larger cross section makes
the system a little hotter. The reason is that there are more
frequent nucleon-nucleon collisions as $\sigma_{NN}$ becomes
larger, which makes  the translation from the longitudinal energy
to thermal energy more efficiently.

%%%%%%%%%%%%%%%%%%%%%%%%%%%%%%%%%%%%%%%%%%%%%%%%%%%%%%%%%%%%%%%%%%%%%
%%%%%%%%%%%%%%%%%%%%%%%%%%%%%%%%%%%%%%%%%%%%%%%%%%%%%%%%%%%%%%%%%%%%%%%

\begin{figure}
\includegraphics[width=9cm]{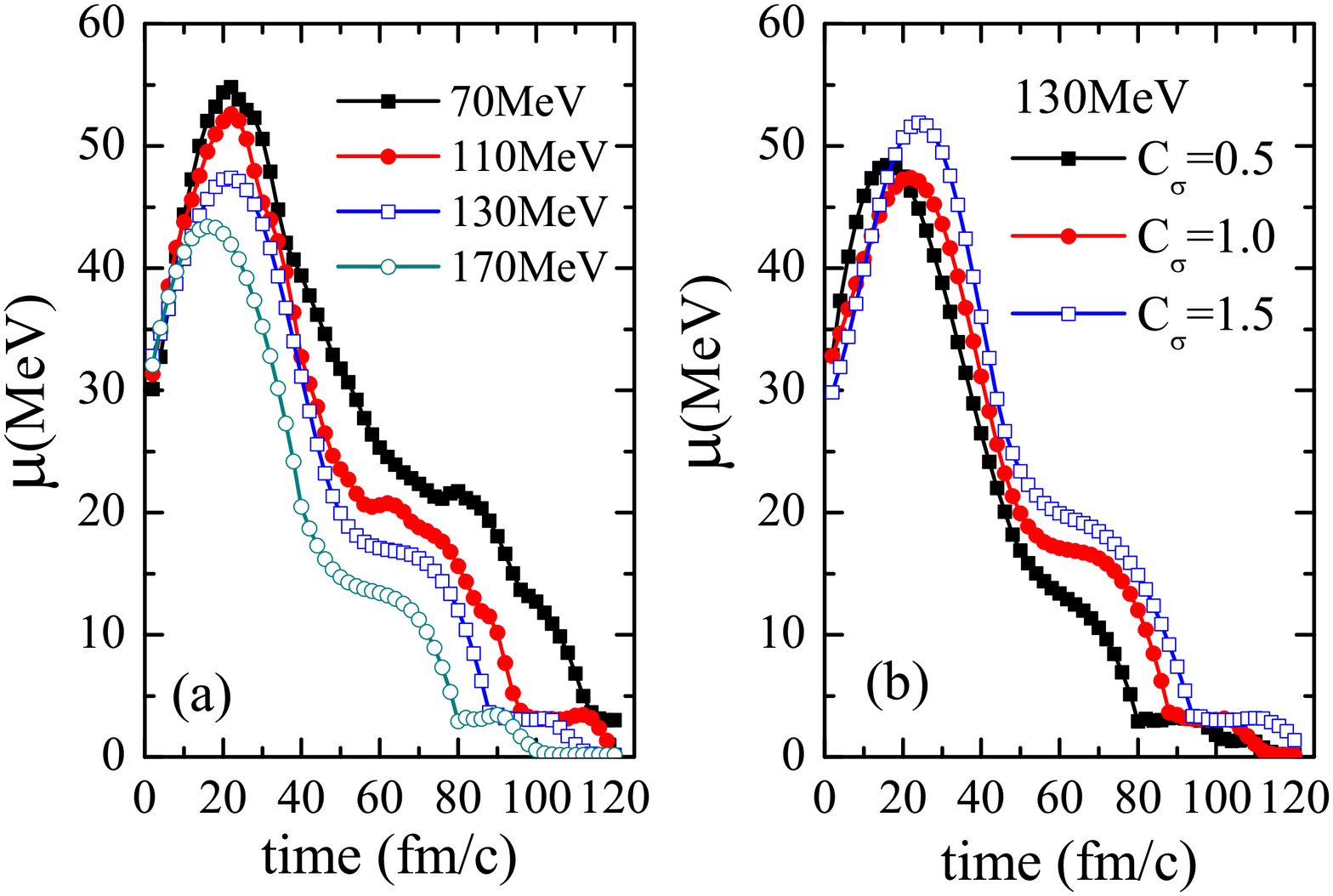}
\vspace{-0.1truein} \caption{\footnotesize (Color online) Same as Fig.~\ref{Fig_temper-merge} but for chemical potential.
}\label{Fig_mu-merge}
\end{figure}

Fig.~\ref{Fig_mu-merge} shows the time evolution of chemical
potential ($\mu$).  Again, the left panel displays the normal
cross section one, we can find that $\mu$ increases in the
compression stage and decreases in the expansion stage, and the
lower the incident energy, the larger the chemical potential.
This might be understood as a large compound nucleus is
formed during the compression stage, and the lower the incident
energy, the larger the compound nucleus is. In panel (b), it shows
that the chemical potential becomes generally larger when
nucleon-nucleon cross section is larger.

\begin{figure}
\includegraphics[width=9cm]{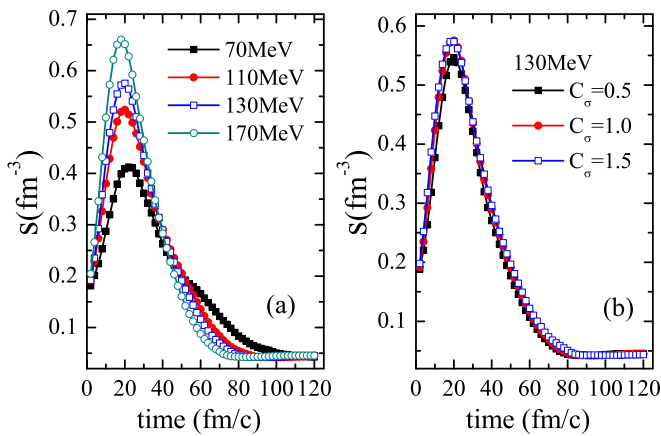}
\vspace{-0.1truein} \caption{\footnotesize (Color online)  Same as Fig.~\ref{Fig_temper-merge} but for
entropy density.
}\label{Fig_sig-merge}
\end{figure}

Time evolution of entropy density is plotted in
Fig.~\ref{Fig_sig-merge}.
It is found that the entropy density almost synchronically evolves with the
temperature. The higher the incident energy and nucleon-nucleon cross section, the larger the entropy density is.

\subsection{Ratio of shear viscosity to entropy density}

Now we can move to the discussion on transport coefficient.  Since
the nuclear participant in central region could be seen as
nuclear fluid, we adopt the Eq.~\ref{Dani-eqn} to calculate the
shear viscosity. Unlike the Green-Kubo formula \cite{kubo,Li2011},
the advantage of the equation is that we can investigate the time
evolution of shear viscosity in the framework of  transport model.
But it should be noted that Eq.~\ref{Dani-eqn} is principally
applicable when the system is largely equilibrated. However, a full
equilibrium is hardly achieved during the whole heavy-ion collision process. So the
shear viscosity extracted here  should be seen as the properties
of an equilibrated nuclear fireball with the same thermodynamic
state as the simulated one.

\begin{figure}
\includegraphics[width=9cm]{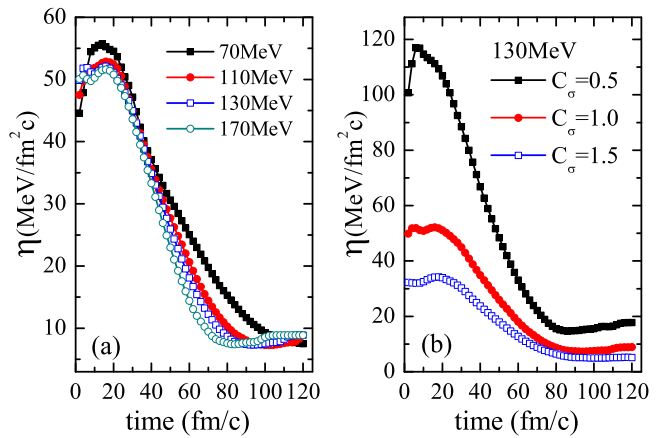}
\vspace{-0.1truein} \caption{\footnotesize (Color online) Same as Fig.~\ref{Fig_temper-merge} but for  shear viscosity $\eta$ of the central fireball.
}\label{Fig_eta-time}
\end{figure}

Fig.~\ref{Fig_eta-time} displays the time evolution of shear
viscosity ($\eta$), it shows an increase in earlier stage and
then drops with time. As Eq.~\ref{Fig_eta-eqn} shows, here the
shear viscosity depends on both temperature and density which
vary with time.  Roughly speaking, the shear viscosity increases
in the compression stage and decreases as the system expands. The
smaller the nucleon-nucleon cross section, the larger the
viscosity in the maximum compression stage. Fig.~\ref{Fig_eta-time}(b) shows that there is a  big
enhancement when the N-N cross section is scaled by $C_{\sigma}=0.5$, which is demonstrated in Eq.\ref{Dani-eqn-cross}, the smaller the N-N cross section the larger the viscosity is.

\begin{figure}
\includegraphics[width=9cm]{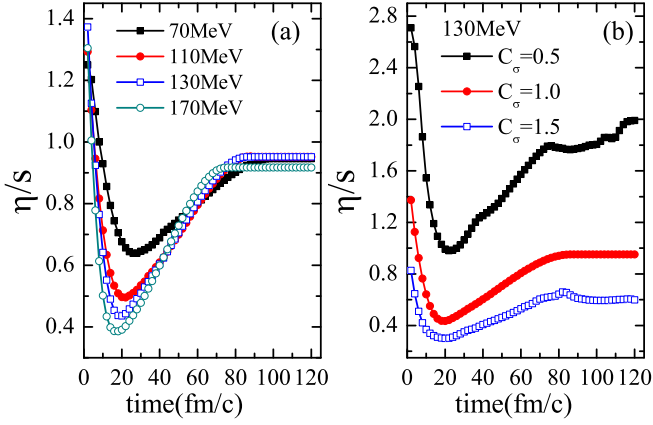}
\vspace{-0.1truein} \caption{\footnotesize (Color online) Same as Fig.~\ref{Fig_temper-merge} but for the ratio of  shear viscosity to entropy density $\eta/s$ of the  central nuclear fireball.
}\label{Fig_ets-merge}
\end{figure}

When the entropy density is taken into account, the ratio of shear
viscosity  to entropy density shows a minimum near maximum
compression point as shown in Fig.~\ref{Fig_ets-merge}. From the hydrodynamical point of view, the less
the $\eta/s$, the more perfect the matter looks like. In this sense, the
nuclear matter becomes a more ideal-like liquid around the most compressible point
in comparison with other evolution stages.  But note that this minimum $\eta/s$ is just
a  transient process. In addition, the
extent of approaching an ideal-like liquid of the nuclear matter
is growing up with the increasing of beam energy.  In relativistic energy
domain, the $\eta/s$ of quark-gluon  matter  becomes very small, close
to 1/4$\pi$ (KSS bound), it is called a perfect liquid.
%%%%%%%%%%%%%%%%%%%%%%%%%%%%%%%%%%%%%%%%%%%
%%%%%%%%%%%%%%%%%%%%%%%%%%%%%%%%%%%%%%%%%%

\begin{figure}
\includegraphics[width=9cm]{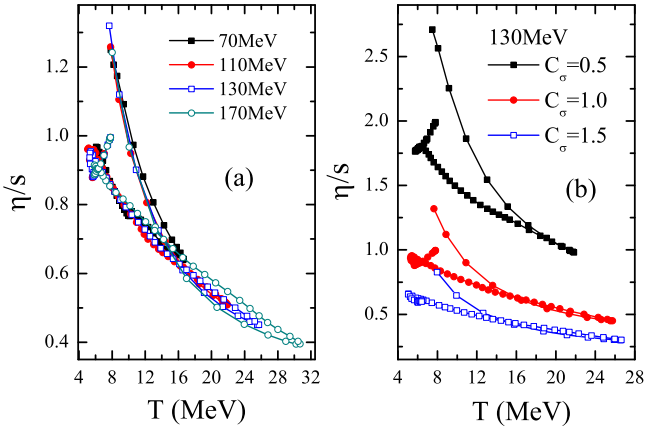}
\vspace{-0.1truein} \caption{\footnotesize (Color online)  (a): The correlation between $\eta/s$ and temperature at different beam energies with a normal nucleon-nucleon cross section parametrization;  (b): the $\eta/s$ evolves  versus temperature at 130 MeV/nucleon with the different nucleon-nucleon cross section. The incident energy and nucleon-nucleon cross section is illustrated in the inset of (a) and (b), respectively.
}\label{Fig_ets-tc}
\end{figure}

Temperature dependence of $\eta/s$ is an important issue to
understand the transport properties of the nuclear matter in
different hot and dense environment.  To this end, we plot a
correlation between $\eta/s$ and temperature in
Fig.~\ref{Fig_ets-tc}(a) at different energies. Note that the density is not fixed
in each curve. It is found that
there is a decrease of $\eta/s$ at first when the system is in the
compression stage. However, $\eta/s$ becomes increasing as the system begins to expand.
The higher the beam energy, the hotter the nuclear matter, and the smaller the $\eta/s$. From this picture it
is obvious to find the time when the $\eta/s$ approaches its  transient
minimum, essentially corresponds that the nuclear matter reaches the
highest temperature. On the other hand,  in the present beam energy
domain below 400 MeV/nucleon, the  transient minimum
of $\eta/s$ which corresponds to the larger compression stage
is around 0.4, which is about 5 times of KSS bound
(i.e. 1/4$\pi$).

%%%%%%%%%%%%%%%%%%%%%%%%%%%%%%%%%%%%%%%%%%%%
\begin{figure}
\includegraphics[width=9cm]{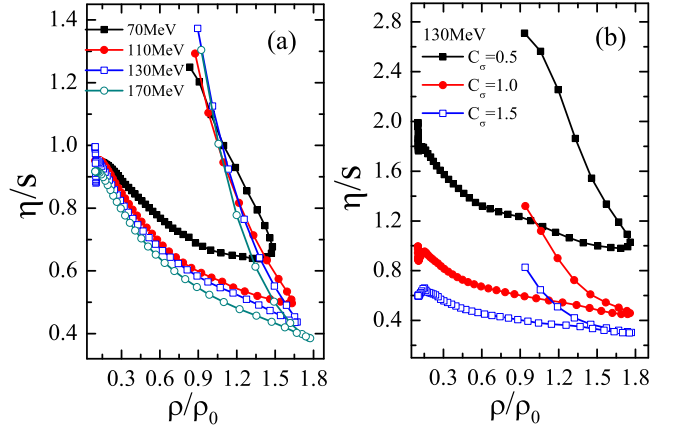}
\vspace{-0.1truein} \caption{\footnotesize (Color online) Same as Fig.~\ref{Fig_ets-tc} but for density dependence of $\eta/s$.
}\label{Fig_ets-rho}
\end{figure}

%%%%%%%%%%%%%%%%%%%%%%%%%%%%%%%%%%%%%%%%%%%%
%%%%%%%%%%%%%%%%%%%%%%%%%%%%%%%%%%%%%%%%%%%%
Furthermore, we can also extract the  correlation between $\eta/s$ and nuclear matter density as shown in Fig.~\ref{Fig_ets-rho}. Here temperature is another hidden variable. Similar to Fig.~\ref{Fig_ets-tc}, $\eta/s$ first drops to a minimum value as the density is compressed to a maximum point and then rises up when the system expands. Larger compressible state produces a less $\eta/s$, i.e. the system is close to a more ideal-like state.

Considering that  only final reaction products can be detected in
experiments,  such as the multiplicity and flows of the fragments
and light particles, it is therefore necessary to check the
$\eta/s$ in the freeze-out stage and see if it is a useful probe to
study the properties of nuclear matter as well as   liquid gas
phase transition. The freeze-out volume has been already studied
in some previous works ~\cite{piantelli2005,parlog2005},
but in our case, it is more suitable to define a freeze-out density instead.
 The time average values of $\eta/s$ when the nuclear matter is in some given
 freeze-out density regions of $\rho/\rho_{0}$ in $[0.19,0.21] , [0.24,0.26]$
 and  $[0.29,0.31]$ have  been extracted as a function of  temperature.
\begin{figure}
\includegraphics[width=9cm]{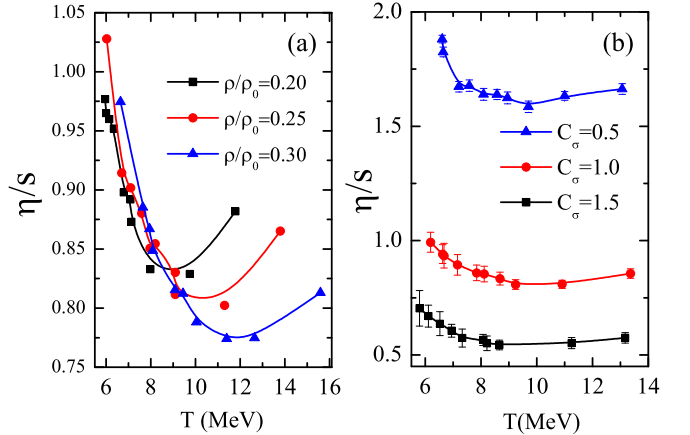}
\vspace{-0.1truein} \caption{\footnotesize (Color online) The average of $\eta/s$  as a function of temperature at different  fixed freeze-out densities (a) and   different cross sections (b).
}\label{Fig_ets-energyn}
\end{figure}

Fig.~\ref{Fig_ets-energyn} shows a correlation of the above
average $\eta/s$  versus temperature for given freeze-out
densities (a) and with different $\sigma_{NN}$ (b). From
Fig.~\ref{Fig_ets-energyn}(a)  we observe  that there exhibits a
local minimum of $\eta/s$ with a value of 0.76 to 0.84 (about 9-10 times of KSS bound),  depending on the
freeze-out density, in the range of 8 - 12 MeV of temperature,
this  phenomenon shall be related to the liquid gas phase
transition. With the increasing of freeze-out density, we observe
the minimal value of $\eta/s$ decreases and while its
corresponding turning temperature increases. The
former is consistent with the results showed in Fig. 12 and the
latter can be understood by the transition temperature/pressure
increases with the freeze-out density as expected by the
pressure-density phase diagram \cite{Gupta,Ma-epja}.
In contrast with the sensitivity of $\eta/s$ to freeze-out
density, Fig.~\ref{Fig_ets-energyn}(b) demonstrates that
the time averaged $\eta/s$ when the system is in a given
$\rho/\rho_{0}$ $[0.2,0.3]$ is also very sensitive to the $\sigma_{NN}$. The larger the N-N cross section is , the smaller the $\eta/s$ is, which means the nuclear matter behaves much more similar as an ideal fluid.

In order to check the result of $\eta/s$, another signal of liquid gas phase transition, namely  intermediate mass fragment, is also studied. The intermediate mass fragment which is  defined  as charge number $Z\in[3,Z_{total}/3]$, where $Z_{total}$ is the total charge number. These fragments are larger than typical evaporated light particles and smaller than the residues and fission products, and they can be considered as nuclear fog. So the multiplicity of intermediate mass fragments ($M_{IMFs}$) is intimately related with the occurrence of liquid gas phase transition. Usually the $M_{IMFs}$  increases first as the collision system changes toward gas phase, and reaches a maximum, then decreases when the system becomes vaporized \cite{Ma1995}.

\begin{figure}
\includegraphics[width=9cm]{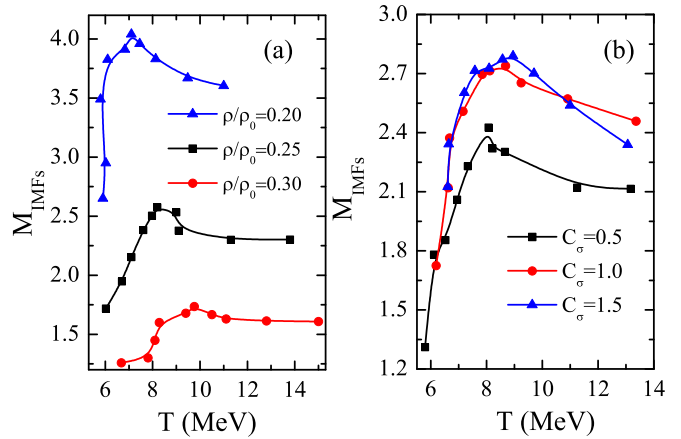}
\vspace{-0.1truein} \caption{\footnotesize (Color online) The average of $M_{IMFs}$  as a function of temperature at different fixed freeze-out density stages (a); the average $M_{IMFs}$ on the whole freeze out stage with $\rho/\rho_{0}\in [0.2,0.3]$ as a function of temperature at different cross sections at the (b).
}\label{Fig_IMF-end}
\end{figure}

The result of $M_{IMFs}$ as a function of temperature is showed in Fig.~\ref{Fig_IMF-end}. In panel (a), it is interesting to find that the higher the density, the lower the transition temperature is, where the maximum of $M_{IMFs}$ approaches. This trend is just coincident with the result of $\eta/s$, which showed in Fig.~\ref{Fig_ets-energyn}(a) except the exact value of the transition temperature; In $M_{IMFs}$ case, the transition temperature $T\in[7,10]$, a little smaller than the $\eta/s$'s, where $T\in[8,12]$. The difference could be explained as the nuclear matter is hotter in the central region, so we argue that the minimum of the $\eta/s$ could be a probe of liquid gas phase transition. In panel (b) the average value of $M_{IMFs}$ as a function of the average temperature when the nuclear matter density $\rho\in [0.2,0.3]\rho_{0}$, it is found that the multiplicities of the intermediate mass fragment increase as the N-N cross section is large. This can be  understood as large N-N cross section increases the  probability of intermediate mass clusters formation. Furthermore, the transition temperature is also increase as the N-N cross section just like the results of Fig.~\ref{Fig_ets-energyn}(b).

It is interesting to note that phase transition temperature $8-12$ MeV which corresponds a
local minimum of $\eta/s$ is basically coincident with previous works \cite{a1,a2}.

\section{\label{sec:level4}Summary and Outlook}

Thermodynamical and transport properties of a  fireball  formed in head-on Au+Au
collisions  are
investigated  in a framework of quantum molecular dynamics model. The
relationships between different thermodynamic quantities are explored. The influences of nucleon-nucleon cross section and
symmetry energy on the thermodynamical and transport properties are also
focused. We found that all the properties are very sensitive to
the nucleon-nucleon cross section and insensitive to the symmetry
energy. In our
calculations, the shear viscosity is calculated by  a
parametrization formula developed by Danielewicz and entropy
density is obtained by a generalized hot Thomas Fermi formalism.
The present work gives a time evolution
of shear viscosity over entropy density ratio of nuclear fireball, which shows that
a transient minimal $\eta/s$ occurs in the largest compression stage. The results at different beam energies
show that the larger the compression, the more ideal the
nuclear fireball behaves like fluid. In the present beam energy domain below 400 MeV/nucleon,
this transient $\eta/s$ approaches to 5 times KSS bound.

In addition,  temperature and density
dependencies of $\eta/s$ are also investigated.
It is of very interesting to observe that a local $\eta/s$ minimum, which is about 9-20 times  KSS bound, emerges
from the temperature dependence of $\eta/s$ at  different constant
freeze-out densities (0.2 - 0.3 $\rho_0$), which corresponds to a liquid-gas phase
transition occurring in the intermediate energy heavy-ion
collisions.And the larger the N-N cross section, the smaller the $\eta/s$ is, which means the nuclear matter behaves more like the ideal fluid. From the temperature dependence of $\eta/s$, we learn
that the phase transition temperature rises up with the freeze-out
density.  In order to check the result of $\eta/s$, another liquid gas phase transition signal, the multiplicity of the intermediate mass fragment is also checked, and a very nice coincidence is found.

Finally, we like to point out that the present work is still  in a  phenomenological level for investigating
$\eta/s$ of hot nuclear matter which is formed in intermediate energy heavy-ion collisions, experimental measurements of  $\eta/s$ are still not available so far.
Therefore, proposals for  direct probes of shear viscosity and entropy density in intermediate energy HIC are very crucial and welcome for constraining the transport properties of nuclear matter around the liquid-gas phase transition.

\section*{ACKNOWLEDGMENTS}
This work is partially supported by the NSFC under contracts No.
11035009, 11220101005, 10979074, 11175231, %10875160, 10805067 and 10975174,
the Major State Basic Research Development Program in China under
Contract No. 2013CB834405,  and  the Knowledge Innovation Project of Chinese
Academy of Sciences under Grant No. KJCX2-EW-N01.

%\end{CJK*}

\end{document}